\documentclass[reprint,aps,prl,twocolumn,superscriptaddress,showpacs]{revtex4-1}
\usepackage[dvips]{graphicx}
\usepackage{color}

\begin{document}

\title{Quantum Critical Scaling for a Heisenberg Spin-$\mathbf{1/2}$ Chain around Saturation}

\author{M. Jeong}
\email{minki.jeong@gmail.com}
\affiliation{Laboratory for Quantum Magnetism, Institute of Condensed Matter Physics (ICMP), Ecole Polytechnique F\'{e}derale de Lausanne (EPFL), CH-1015 Lausanne, Switzerland}
\author{H. M. R{\o}nnow}
\affiliation{Laboratory for Quantum Magnetism, Institute of Condensed Matter Physics (ICMP), Ecole Polytechnique F\'{e}derale de Lausanne (EPFL), CH-1015 Lausanne, Switzerland}

\begin{abstract}
We demonstrate quantum critical scaling for an $S=1/2$ Heisenberg antiferromagnetic chain compound CuPzN in a magnetic field around saturation, by analysing previously reported magnetization [Y. Kono {\it et al.}, Phys. Rev. Lett. {\bf 114}, 037202 (2015)], thermal expansion [J. Rohrkamp {\it et al.}, J. Phys.: Conf. Ser. {\bf 200}, 012169 (2010)] and NMR relaxation data [H. K\"uhne {\it et al.}, Phys. Rev. B {\bf 80}, 045110 (2009)]. The scaling of magnetization is demonstrated through collapsing the data for a range of both temperature and field onto a single curve without making any assumption for a theoretical form. The data collapse is subsequently shown to closely follow the theoretically-predicted scaling function without any adjustable parameters. Experimental boundaries for the quantum critical region could be drawn from the variable range beyond which the scaled data deviate from the theoretical function. Similarly to the magnetization, quantum critical scaling of the thermal expansion is also demonstrated. Further, the spin dynamics probed via NMR relaxation rate $1/T_1$ close to the saturation is shown to follow the theoretically-predicted quantum critical behavior as $1/T_1\propto T^{-0.5}$ persisting up to temperatures as high as $k_\mathrm{B}T \simeq J$, where $J$ is the exchange coupling constant.
\end{abstract}

\maketitle

\begin{figure}
\includegraphics[width=0.45\textwidth]{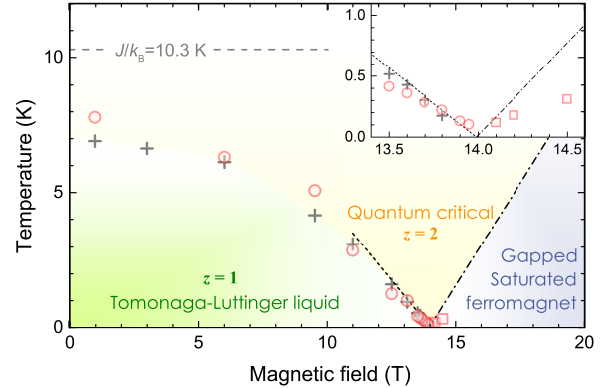}
\caption{(Color online) Phase diagram of CuPzN where crosses represent the magnetization $M(T)$ maxima as reported in Ref.~\cite{Kono15PRL}. Dotted and dash-dotted lines steming from the saturation field represent the theoretical $M(T)$ maximum and the gap size, respectively. Squares and circles represent the proposed experimental boundaries of quantum critical region (see the text). Inset shows a zoom in close to the saturation field.}
\label{fig1}
\end{figure}

A quantum critical point (QCP) is a zero-temperature singularity in the phase diagram of matter forming the border between two competing ground states \cite{Sachdev11PT}. It is driven by a non-thermal parameter such as a magnetic field, pressure, or chemical substitution, and characterized by strong quantum fluctuations. While a QCP is defined strictly at zero temperature, the interplay between quantum and thermal fluctuations gives rise to a so-called quantum critical region at finite temperatures in an extended parameter space (illustrated by the yellow fan-out area in Fig.~\ref{fig1}). This intriguing region is characterized by the absence of energy scales other than temperature as well as the corresponding critical properties of physical observables, e.g. correlation or response functions, which culminate into scaling behavior and universality \cite{Sachdev11PT, Sachdev, Sondhi97RMP, Vojta03RPP}. Such quantum criticality has been experimentally observed or inferred in diverse systems including magnetic insulators \cite{Lake05NatMat, Merchant14NatP, Kinross14PRX}, organic conductors \cite{Kagawa05Nat}, heavy fermions \cite{Gegenwart08NatP, Stockert11ARCMP}, cuprates \cite{Broun08NatP}, pnictides \cite{Shibauchi13ARCMP}, and cold atoms \cite{DonnerSci07}, and is widely believed to underpin exotic phenomena like unconventional superconductivity. However, understanding quantum criticality through connecting microscopics to experimental observation largely remains challenging \cite{Sachdev11PT, Kagawa05Nat, Gegenwart08NatP, Stockert11ARCMP, Broun08NatP, Shibauchi13ARCMP, Coleman05Nat}.

Quantum magnets are an ideal playground in that respect owing to their simple and well-defined Hamiltonian \cite{Sachdev08NatP}. In particular, one-dimensional (1D) spin systems for which exact solutions are available may serve a testbed for quantitative comparison between theories and experiments \cite{Giamarchi, Klanjsek08PRL, Mourigal13NatP, Schmidiger13PRL}. Indeed, quite a few excellent quasi-1D quantum magnets having accessible critical field strength, i.e. relatively small exchange coupling strength, have been synthesized in single crystals \cite{Landee13EJIC, Yankova12PM}, which triggered activities for experimentally probing various field-induced quantum criticality \cite{Watson01PRL, Mukhopadhyay12PRL, Povarov15PRB, Kono15PRL}.

\begin{figure*}
\includegraphics[width=0.85\textwidth]{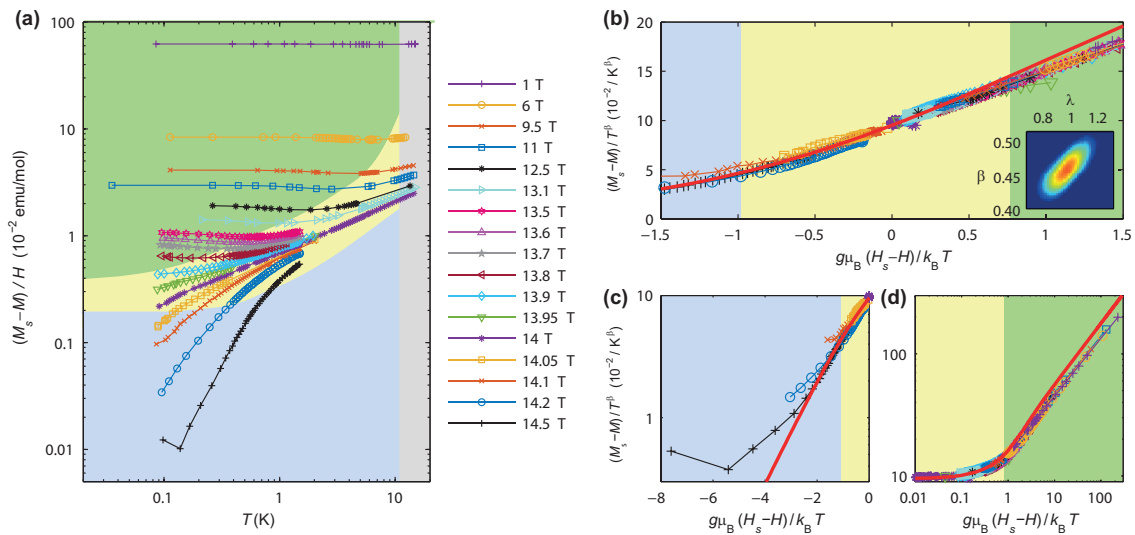}
\caption{(Color online) (a) $(M_s-M)/H$ as a function of temperature at different fields, taken from Ref.~\cite{Kono15PRL}. The data on blue and green background belong to ferromagnetic and TLL phases, respectively. The data set on a yellow background are used for data collapse and scaling analysis. (b) The data collapse when $M_s-M$ scaled by $T^\beta$ is plotted against a variable $g\mu_\mathrm{B}(H_s-H)/k_\mathrm{B}T^\lambda$ with $\lambda=1$, where the parameters were obtained by fitting the collapsed data to a third-order polynomial. Inset shows a slice of the colormap of goodness-of-fit at the best-fit $H_s$ as a function of $\beta$ and $\lambda$. Solid line is the theoretical scaling function (Eq.~\ref{eq:scaling} and \ref{eq:funcM} in the text). (c) The data collapse for $H>H_s$ and (d) for $H<H_s$}
\label{fig2}
\end{figure*}

Arguably the simplest model to capture quantum criticality, a nearest-neighbor $S=1/2$ Heisenberg antiferromagnetic chain, is realized in the organometallic compound $\mathrm{Cu(C_4H_4N_2)(NO_3)_2}$, CuPzN for short \cite{Santoro70ACSB, Hammar99PRB}. This material has a relatively small exchange constant $J/k_\mathrm{B}=10.3$ K along the chain direction (crystallographic $a$ axis of an orthorhombic structure) \cite{Hammar99PRB}, which results in a laboratory-accessible saturation field $H_s=2J/g\mu_\mathrm{B}\simeq 13.9-15$ T depending on the field orientation \cite{gfactor, McGregor76JCP, Validov14JPCM}. Its ground state is a Tomonaga-Luttinger liquid (TLL) for $H<H_s$ and a saturated ferromagnet with a gap for $H>H_s$, leaving a QCP at $H=H_s$ (see Fig.~\ref{fig1}). The field-induced quantum criticality of CuPzN has been studied by using various experimental techniques including magnetization \cite{Kono15PRL}, thermal expansion \cite{Rohrkamp09JPCS}, and thermal transport \cite{Sologubenko07PRL}  measurements. The most recent high-precision magnetization measurements \cite{Kono15PRL}, for instance, demonstrated a theoretically predicted power-law behavior at $H_s$, $(M_s-M) \propto T^{\beta}$, where $M_s$ is the saturated magnetization, finding critical exponent $\beta=0.48(1)$ in excellent agreement with the theoretical $\beta=0.5$ \cite{Affleck91PRB, Sachdev94PRB}.

However, despite the high quality data being available, we find that for this simple spin-chain model the most dramatic manifestation of quantum criticality, namely, \emph{quantum critical scaling}, has not been explored. Such scaling is a direct consequence of the absence of energy scale other than temperature such that the measured quantities, when scaled by the temperature to a certain universal power, collapse onto a single curve for the plot against an appropriate scaling variable \cite{Sachdev, Sondhi97RMP, Vojta03RPP}. To fully assess the universality of a quantum critical region requires a demonstration of this scaling behavior.

Here we analyze the reported magnetization \cite{Kono15PRL} and thermal expansion \cite{Rohrkamp09JPCS} data, and successfully demonstrate excellent scaling behavior over a wide range of temperature and field around $H_s$. The collapsed magnetization data closely follow the theoretical scaling function for a certain range, which allows us to draw experimental boundaries of the quantum critical region. Subsequently, we revisit the reported NMR relaxation data \cite{Kuhne09PRB} to show that the spin dynamics close to $H_s$ display the theoretically predicted power-law behavior for a quantum critical region, up to rather high temperatures $k_\mathrm{B}T \simeq J$.

Figure~\ref{fig2}(a) shows $(M_s-M)/H$ as a function of temperature in different fields $H\parallel b$, extracted from Ref.~\cite{Kono15PRL} where they reported $H_s=13.97(6)$ T. To test for quantum critical scaling, the data should be filtered such that those belonging to the neighboring TLL and ferromagnetic phases are excluded. The crossover temperature $T^*$ separating a TLL at low temperatures from a high-temperature phase is marked by a cusp-like maximum in $M(T)$ for $H$ close to $H_s$ \cite{Maeda07PRL}, which smoothly connects to a broad $M(T)$ maximum at lower fields characteristic of the development of antiferromagnetic correlations in a spin chain \cite{Muller81PRB}. These $M(T)$ maxima (crosses in Fig.~\ref{fig1}), close to $H_s$ in particular, were shown to follow closely the theoretical prediction $k_\mathrm{B}T^* = 0.76328\;g\mu_\mathrm{B}(H_s-H)$ \cite{Maeda07PRL, Kono15PRL} presented by a dotted line in Fig.~\ref{fig1}. This leads us to use for $H<H_s$ only the data belonging to $T>T^*$. On the other side of the phase diagram, the gap size $\Delta$ in the ferromagnetic phase scales linearly with the field-difference from $H_s$, i.e. $\Delta = g\mu_\mathrm{B}(H-H_s)$ \cite{Muller81PRB}, as presented by dash-dotted line in Fig.~\ref{fig1}. We therefore use for $H>H_s$ only the data belonging to $k_\mathrm{B}T>\Delta$. Lastly, we used only the data for $k_\mathrm{B}T<J$ to exclude thermal paramagnetic phase (grey background in Fig. \ref{fig2}(a)). The data set on a yellow background in Fig.~\ref{fig2}(a) corresponds to the filtered ones using the reported $H_s$ \cite{Kono15PRL}.

We perform a data collapse without making any assumption for a theoretical scaling function. A minimal form is assumed as 
\begin{equation}
M_s-M=T^\beta\Phi\left(\frac{g\mu_\mathrm{B}(H_s-H)}{k_\mathrm{B}T^\lambda}\right).
\end{equation} 
The parameters $\beta$, $\lambda$, and $H_s$ are determined by fitting the data to an arbitrary function $\Phi$ represented by a third-order polynomial with free parameters. We obtained as best-fit $\lambda=0.98(2)$, which indicates that $g\mu_\mathrm{B}(H_s-H)$ linearly scales with temperature. This is a signature of underlying quantum criticality \cite{Sachdev, Sachdev94PRB}. The obtained best-fit $\beta=0.465(5)$ and $H_s=14.00(3)$ T also agree with the temperature exponent for $M_s-M$ at $H_s$ and the reported $H_s$ value, respectively \cite{Kono15PRL}. The inset of Fig.~\ref{fig2} shows a slice of the colormap of goodness-of-fit, defined by the residual-sum-of-squares divided by the number of degrees of freedom, as a function of $\beta$ and $\lambda$ at the best-fit $H_s$. Figure~\ref{fig2}(b) plots $(M_s-M)/T^\beta$ against $g\mu_\mathrm{B}(H_s-H)/k_\mathrm{B}T^\lambda$ using the best-fit parameters while setting $\lambda=1$. We find that the data accurately collapse onto a single curve highlighting the scaling behavior.  Moreover, the data collapse is found to persist far beyond the fit range, as shown in Fig.~\ref{fig2}(c) and (d) for the $H>H_s$ and $H<H_s$ ranges, respectively.

\begin{figure}
\includegraphics[width=0.4\textwidth]{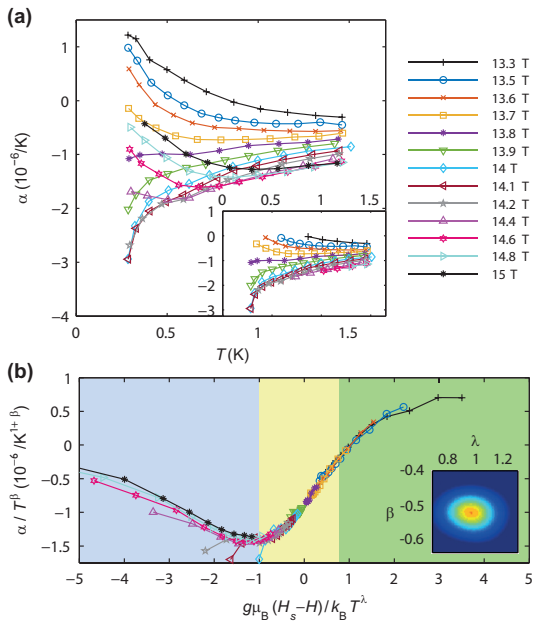}
\caption{(Color online) (a) Thermal expansion $\alpha$ as a function of temperature at different fields, taken from Ref.~\cite{Rohrkamp09JPCS}. Inset displays the filtered data set used for the data collapse and scaling analysis. (b) The data collapse by plotting $\alpha$ scaled by $T^{\beta}$ against a variable $g\mu_\mathrm{B}(H_s-H)/k_\mathrm{B}T^\lambda$. The parameters $\beta$ and $\lambda$ as well as $H_s$ were obtained from the best fit of the collapsed data to a third-order polynomal. Inset shows a slice of the colormap of goodness-of-fit at the best $H_s$ as a function of $\beta$ and $\lambda$.}
\label{fig3}
\end{figure}

Next we compare the data collapse to the existing theory \cite{Sachdev94PRB}. The magnetization close to a field-induced QCP with the dynamical exponent $z=2$, i.e. quadratic dispersion, for a Heisenberg antiferromagnet of dimension $d<2$ is predicted to follow the scaling form 
\begin{equation}\label{eq:scaling}
M_s - M = g\mu_\mathrm{B}\left(\frac{2k_\mathrm{B}T}{J}\right)^{\beta}\mathcal{M}\left(\mu/k_\mathrm{B}T\right),
\end{equation} 
where $\mu\equiv g\mu_\mathrm{B}(H_s-H)$ plays the role of chemical potential and the exponent $\beta=d/2$ \cite{Sachdev94PRB}. In a 1D dilute magnon limit close to $H_s$, mapping of interacting magnons onto free fermions leads to the expression \cite{Affleck91PRB, Sachdev94PRB},
\begin{equation}\label{eq:funcM}
\mathcal{M} = \frac{1}{\pi}\int_0^\infty\frac{1}{\mathrm{e}^{x^2-\mu/k_\mathrm{B}T}+1}\mathrm{d}x.
\end{equation} 
The solid line in Figs.~\ref{fig2}(b-d) represents the theoretical scaling function Eq.~\ref{eq:funcM}. We find that the collapsed data almost perfectly follow the theoretical function with \emph{no adjustable parameters}, which is a clear experimental confirmation \cite{Watson} of the hypothesis coined as the zero scale-factor universality \cite{Sachdev94PRB}.

While the data collapse extends well beyond the fit range, the data begin to show gradual deviation from the theoretical function when moved sufficiently away from $H_s$. This is shown in Fig.~\ref{fig2}(c) and (d) for $H>H_s$ and $H<H_s$, respectively. The deviation for $\mu/k_\mathrm{B}T\ll -1$ (Fig.~\ref{fig2}(c)) reflects the opening of the gap by entering a nonuniversal, ferromagnetic phase. On the other hand, the collapsed data for $H<H_s$ (Fig.~\ref{fig2}(d)) still accurately fall on a single curve up to the highest measured $\mu/k_\mathrm{B}T\sim 3\times 10^2$ despite a systematic departure from the theoretical curve for $\mu \gtrsim k_\mathrm{B}T$. The data for $\mu/k_\mathrm{B}T\gg 1$ belong to the TLL which itself is genuinely a quantum-critical state. However, TLL belongs to a different universality class having a linear dispersion, i.e. $z=1$, of which quantum criticality including scaling behavior has been widely investigated \cite{Lake05NatMat, Povarov15PRB, Dender97, Stone03PRL, Klanjsek08PRL, Jeong13PRL}. By locating the variable range beyond which the data begins to show deviation from the theoretical curve by 10 \%, for instance, we can draw experimental boundaries of quantum critical region as shown in Fig.~\ref{fig1} by squares for $H>H_s$ and circles for $H<H_s$. It may be noteworthy that the scaling behavior for $H>H_s$ persists down to lower temperatures far below the gap size.

We further test quantum critical scaling by examining the thermal expansion $\alpha$. Figure \ref{fig3}(a) shows $\alpha(T)$ at different $H\parallel b$, taken from Ref.~\cite{Rohrkamp09JPCS}, where the inset shows the data set filtered according to the same criteria as for the magnetization. Figure~\ref{fig3}(b) shows $\alpha$ scaled by $T^\beta$ against $g\mu_\mathrm{B}(H_s-H)/k_\mathrm{B}T^\lambda$ where $\beta$, $\lambda$, and $H_s$ were obtained by fitting the collapsed data to a thrid-order polynomial. Again, the data collapse is excellent with the best-fit $\lambda=0.97(15)\simeq 1$ witnessing quantum criticality, the $\beta=-0.53(8)$ in agreement with the theoretical $-0.5$ \cite{Zhu03PRL}, and $H_s=13.89(9)$ T being consistent with the reported value \cite{Kono15PRL, Rohrkamp09JPCS}. Similarly to the magnetization, the data collapse extends far beyond the fit range (yellow background).

\begin{figure}
\includegraphics[width=0.4\textwidth]{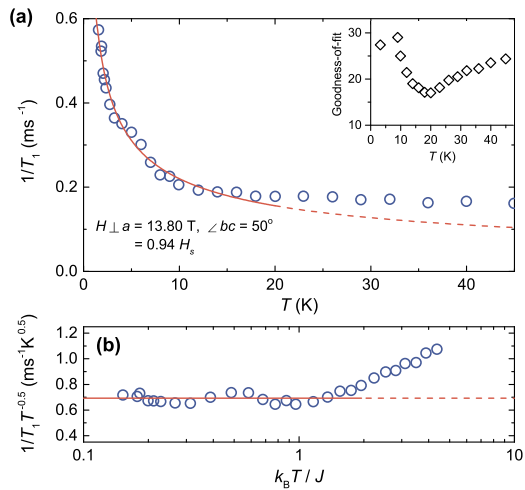}
\caption{(Color online) (a) $^{13}$C NMR relaxation rate $1/T_1$ as a function of temperature at $H=0.94H_s$, taken from Ref.~\cite{Kuhne09PRB}. Solid line is the best fit, $\theta=-0.50(2)$ for $0<T\leq 20$ K, to the power-law $1/T_1\propto T^\theta$. Inset shows the goodness-of-fit as a function of upper temperature bound for the fit range. (b) A scaled plot of $1/T_1(T)$ divided by $T^{-0.5}$ against $k_\mathrm{B}T/J$.}
\label{fig4}
\end{figure}

Now we turn our attention to the spin dynamics near $H_s$ probed via NMR relaxation rate $1/T_1$ measurements. Figure \ref{fig4}(a) reproduces the temperature dependence of $^{13}$C $1/T_1$ in $H=13.80$ T, taken from Ref.~\cite{Kuhne09PRB}. This field value corresponds to $0.94H_s$ for the given orientation, i.e. $H\perp a$ and $50^\circ$ from $b$ to $c$ \cite{Kuhne09PRB, Kuhne11PRB}. NMR $1/T_1$ probes local electron spin correlations in the low energy limit, and a power-law behavior $1/T_1\propto T^\theta$ is expected for a gapless, quantum-critical region \cite{Sachdev, Giamarchi, Kinross14PRX}. We tried to fit the data to the power law while varying the upper bound on the included temperature range. The inset of Fig.~\ref{fig4}(a) shows the goodness-of-fit of which the minimum is obtained for the $T\leq 20$ K range. The solid line in Fig.~\ref{fig4}(a) is the corresponding best-fit result, which yields $\theta=-0.50(2)$.

Theoretically, scaling arguments for a 1D QCP with $z=2$ leads to $1/T_1\propto T^{-0.5}$ \cite{Orignac07PRB}. Our fit result perfectly agrees with this theoretical form. Meanwhile, the $1/T_1$ data have been originally treated within the framework of TLL with the help of field-theoretic calculations \cite{Kuhne11PRB}. A TLL as a gapless quantum-critical-state supports similarly a power-law $1/T_1$ behavior \cite{Giamarchi, Jeong13PRL, Chitra97PRB, Giamarchi99PRB, Yoshioka02JPCS, Singer05PRL, Dora07PRL, Ihara10EPL, Zhi15PRL}. The corresponding exponent $\theta$ of a TLL of quantum magnets is a function of $M$ (and thus $H$) that effectively controls the spinon interactions \cite{Giamarchi}. When the TLL is tuned toward the limit of noninteracting spinons, i.e. $H\rightarrow H_s$, the exponent approaches a universal value as $\theta\rightarrow -0.5$ \cite{Chitra97PRB, Giamarchi99PRB, Jeong13PRL}. This may leave certain ambiguity whether the observed $\theta=-0.5$ corresponds to 1D QCP with $z=2$ or noninteracting TLL with $z=1$. However, for the given $H=0.94H_s$, the upper bound $T^*$ for the TLL is expected only $\sim 1$ K \cite{Kono15PRL, Maeda07PRL}, whereas the $1/T_1$ data and the power-law fit were obtained for $T>1$ K up to an order of magnitude higher than $T^*$. Thus we suggest that the observed $1/T_1(T)\propto T^{-0.5}$ dictates the $z=2$ QCP.

It was only recently that a fundamental question to what extent quantum criticality would persist up in temperature \cite{Kopp05NatP} was quantitatively addressed in experiments \cite{Kinross14PRX, Klanjsek14Phys}. The $^{93}$Nb $1/T_1$ measurements on a transverse-field quasi-1D Ising ferromagnet $\mathrm{CoNb_2O_6}$ showed quantum critical behavior up to as high a temperature as $0.4$ times the underlying exchange coupling scale \cite{Kinross14PRX}. This could be in line with the present result for CuPzN. Figure~\ref{fig4}(b) plots the $1/T_1(T)$ scaled by $T^{-0.5}$ against the normalized temperature $k_\mathrm{B}T/J$. This plot emphsaizes that the power-law or quantum criticality persists up to a temperature as high as $k_\mathrm{B}T\simeq J$, being consistent with the previous thermodynamic measurements of magnetization and specific heat \cite{Kono15PRL}.

{\it Conclusion.} By revisiting the existing experimental data, we could demonstrate quantum critical scaling and define the quantum critical region for a spin chain compound CuPzN around saturation. The scaling behavior for magnetization and thermal expansion was demonstrated without making an assumption for a theoretical function. The collapsed magnetization data closely follow the theoretical scaling function of the zero scale-factor universality for $z=2$ for an extended variable range, which allows us to draw the experimental boundaries of the quantum critical region. The spin dynamics close to the saturation probed via NMR relaxation display the theoretically predicted power-law behavior characteristic of quantum criticality up to as high a temperature as $k_\mathrm{B}T\simeq J$.

M.J. thanks M. Klanj\v{s}ek for comments on the manuscript. This work was funded by the Swiss National Science Foundation and its Sinergia network MPBH, Marie Curie Action COFUND (EPFL Fellows), and European Research Council grant CONQUEST.


\begin{thebibliography}{99}
\bibitem{Watson}The only similar analysis of the magnetization to our knowledge was reported in Ref.~\cite{Watson01PRL} for the QCP of a strong-rung spin ladder compound BPCB, but the data collapse to the theoretical function was not as clear.
\end{thebibliography}
\end{document}